\documentclass[review,12pt,times]{elsarticle}
\usepackage{epsfig,amsmath,amssymb}

\newcommand*{\TITLE}[1]{\section{#1}}
\newcommand*{\SUBTITLE}[1]{\subsection{#1}}
\newcommand*{\be}[0]{\begin{equation}}
\newcommand*{\ee}[0]{\end{equation}}
\newcommand*{\beu}[0]{\begin{equation*}}
\newcommand*{\eeu}[0]{\end{equation*}}
\newcommand*{\ba}[0]{\begin{array}}
\newcommand*{\ea}[0]{\end{array}}
\newcommand*{\bfig}[0]{\begin{figure}}
\newcommand*{\efig}[0]{\end{figure}}
\newcommand*{\bfigwide}[0]{\begin{figure*}}
\newcommand*{\efigwide}[0]{\end{figure*}}
\newcommand*{\Fig}[1]{Figure~\protect\ref{fig:#1}}
\newcommand*{\Tab}[1]{Table~\protect\ref{tab:#1}}
\newcommand*{\Eqn}[1]{Eqn.~(\protect\ref{eqn:#1})}
\newcommand*{\Sec}[1]{Section~\protect\ref{sec:#1}}
\newcommand*{\et}[0]{\textit{et~al.}}

\newcommand*{\eV}[0]{\text{eV}}
\newcommand*{\meV}[0]{\text{meV}}
\newcommand*{\C}[0]{^\circ\text{C}}
\newcommand*{\kB}[0]{k_{\text{B}}}
\newcommand*{\MPa}[0]{\text{MPa}}
\newcommand*{\x}[0]{\times}
\newcommand*{\EE}[1]{\x 10^{#1}}

\newcommand*{\eps}[0]{\varepsilon}

\newcommand*{\conc}[0]{c_\text{s}}	
\newcommand*{\concmin}[0]{\conc^{\text{min}}}	
\newcommand*{\concopt}[0]{\conc^{0}}	
\newcommand*{\disl}[0]{\rho_\perp}
\newcommand*{\ave}[1]{\overline{#1}}
\newcommand*{\DE}[0]{\Delta E}
\newcommand*{\DEave}[0]{\ave{\DE}}
\newcommand*{\dE}[0]{\delta E}
\newcommand*{\DEabs}[0]{\ave{|\DE|}}
\newcommand*{\dDE}[0]{\Big(\dE^2+\DEave^2\Big)}
\newcommand*{\Pdk}[0]{P_{\text{dk}}}
\newcommand*{\tauac}[0]{\tau^{\text{a}}}
\newcommand*{\tauacmin}[0]{\tau^{\text{a}}_{\text{min}}}
\newcommand*{\taua}[0]{\tau_{\text{ath}}}
\newcommand*{\taus}[0]{\tau^\star}
\newcommand*{\Sc}[0]{S_{\text{c}}}
\newcommand*{\lkink}[0]{\ell_{\text{kink}}}
\newcommand*{\Hf}[0]{E_{\text{f}}}
\newcommand*{\adk}[0]{\alpha_{\text{dk}}}
\newcommand*{\Hdk}[0]{H_{\text{dk}}}
\newcommand*{\Hdkb}[0]{H^0_{\text{dk}}}
\newcommand*{\nuf}[0]{\nu_{\text{dk}}}
\newcommand*{\Gdk}[0]{G^{\text{dk}}}
\newcommand*{\Gdka}[0]{\Gdk_{\text{app}}}
\newcommand*{\Gkm}[0]{G^{\text{km}}}

\newcommand*{\tdk}[0]{t_{\text{dk}}}
\newcommand*{\tkm}[0]{t_{\text{km}}}
\newcommand*{\Htau}[0]{E_\tau}
\newcommand*{\Hkm}[0]{\langle E^2\rangle_{\text{km}}}

\newcommand*{\aaxis}[0]{\frac{a}{3}[2\bar{1}\bar{1}0]}
\newcommand*{\baxis}[0]{a[01\bar{1}0]}
\newcommand*{\caxis}[0]{c[0001]}
\newcommand*{\bv}[0]{\vec{b}}

\newlength{\wholefigwidth}
\newlength{\smallfigwidth}
\newlength{\halfsmallfigwidth}
\journal{Acta Materialia}

\begin{document}
\begin{frontmatter}

\title{Prediction of thermal cross-slip stress in magnesium alloys from direct first principles data}

\author{Joseph A. Yasi}
\address{Department of Physics, University of Illinois at Urbana-Champaign, Urbana, IL 61801}
\author{Louis G. Hector, Jr.}
\address{General Motors R\&D Center, 30500 Mound Road, Warren, MI 48090}
\author{Dallas R. Trinkle}
\address{Department of Materials Science and Engineering, University of Illinois at Urbana-Champaign, Urbana, IL 61801}
\ead{dtrinkle@illinois.edu}

\begin{abstract}
We develop a first-principles model of thermally-activated cross-slip in magnesium in the presence of a random solute distribution.  Electronic structure methods provide data for the interaction of solutes with prismatic dislocation cores and basal dislocation cores.  Direct calculations of interaction energies are possible for solutes---K, Na, and Sc---that lower the Mg prismatic stacking fault energy to improve formability.  To connect to thermally activated cross-slip, we build a statistical model for the distribution of activation energies for double kink nucleation, barriers for kink migration, and roughness of the energy landscape to be overcome by an athermal stress.  These distributions are calculated numerically for a range of concentrations, as well as alternate approximate analytic expressions for the dilute limit.  The analytic distributions provide a simplified model for the maximum cross-slip softening for a solute as a function of temperature.  The direct interaction calculations predict lowered forming temperatures for Mg-0.7at.\%Sc, Mg-0.4at.\%K, and Mg-0.6at.\%Na of approximately 250$^\circ$C.
\end{abstract}


\end{frontmatter}

\TITLE{Introduction}
Increased interest in the light-weight structural metal magnesium\cite{MgTech2006} to replace aluminum or steels in automotive applications\cite{Pollock2010} has focused attention on a variety of metallurgical issues, including formability.  Current Mg alloys require temperatures near $300\C$ for forming to activate the five independent slip systems required by the von~Mises criterion\cite{Taylor1938}; this is in part due to the large anisotropy between basal and prismatic slip\cite{Agnew2005}.  Cross-slip of $a$-type dislocations from the easy $(0001)$ basal plane onto the hard $(01\bar10)$ prismatic plane requires large stresses or high temperatures.  Experimentally, few solutes have been found to lower the stress for cross-slip: Al and Zn lower the stress at low (below room) temperatures\cite{Akhtar1969b}, while Li can lower the cross-slip stress in both regimes\cite{Ahmadieh1965,Urakami1970a,Urakami1970b,Urakami1971}.  The difficulty of performing experiments to measure cross-slip stresses for alloys---requiring single-crystal samples oriented for prismatic slip---is compounded by the possibility that, like solid-solution softening in BCC alloys\cite{Pink1979}, it may occur over a limited concentration and temperature range.  Hence, new state-of-the-art first-principles prediction of solute/dislocation interactions coupled with predictive computational modeling of thermally-activated cross-slip in the presence of solutes is necessary to guide the design of new alloys.

Couret and Caillard \textit{in situ} experimental measurements\cite{Couret1985.1,Couret1985.2} found that at high (above room) temperatures, cross-slip in magnesium is the result of a double-kink nucleation (also called ``jog-pair''\cite{Yoshinaga1963}) mechanism.  A basal screw dislocation constricts at kinks of height $c$ and spreads on two neighboring basal planes\cite{Couret1991}.  At low (below room) temperatures, cross-slip instead occurs by constriction and bowing of a screw dislocation---the Friedel-Escaig mechanism\cite{Friedel1959,Escaig1968}; as we are interested in the problem of forming near $300\C$, we consider only the double-kink nucleation mechanism.  We model double-kink nucleation by computing the geometry and formation energy of a single basal-to-prismatic kink with a validated embedded-atom potential\cite{Sun2006}; together with the Peierls stress, we accurately reproduce the experimental cross-slip stress over the 300K--700K temperature range.  The starting geometry and energetics are combined with first-principles modeling of the changes in energy from basal to prismatic cores due to substitutional solutes in magnesium.  Finally, we compute---numerically and with an analytic approximation---the distribution of double-kink activation barriers and energy barriers for kink migration to predict the stress for thermally-activated cross-slip with solute concentration and temperature for K, Na, and Sc.  In order to do \textit{direct} substitution of solutes, we can only consider solutes which do not increase the energy of the prismatic core.  Finally, we use our model with solute data for softening to predict the maximum possible reduction in forming temperature for a binary alloy.

\TITLE{Computational methodology}
To accurately compute the interaction of solutes with Mg dislocation cores, we use density-functional theory with flexible boundary conditions for a stress-free dislocation core.  Density functional theory calculations are performed with \textsc{vasp}\cite{Kresse93,Kresse96b}, using a plane-wave basis with the projector augmented-wave (PAW) method\cite{Blochl1994} and potentials generated by Kresse\cite{Kresse1999}.  The many-electron exchange and correlation effects are described by a generalized gradient approximation\cite{Perdew92}, and a plane-wave kinetic energy cutoff of 273eV ensures accurate treatment of the potentials.  The electronic configurations of the PAW potentials are Mg: [Ne]$3s^2$, K: [Mg]$3p^64s^1$, Na: [Ne]$3s^1$, Sc: [Ar]$4s^13d^2$.  The dislocation calculation use a $1\x1\x16$ $k$-point mesh with a Methfessel-Paxton smearing of 0.5eV to give an energy accuracy of 5meV for bulk Mg.  The dislocation geometry and flexible boundary condition method\cite{Sinclair1978,Rao1998,TrinkleLGF2008} use a similar geometry to previous calculations for basal dislocations\cite{Yasi2010} with a reshaped region I for prismatic splitting; we relax atomic positions until the forces are less than 5meV/\AA, where the lattice Green function is used to displace atoms in the outer regions.  To produce a prismatic screw dislocation, we begin relaxation from the anisotropic displacement field from two $b/2$ partial screw dislocations (for total Burgers vector $b=\aaxis$) separated by $c$ in the prismatic plane.

Determining the geometry and formation energy for a basal to prismatic kink requires a validated classical potential\cite{Sun2006}.  This EAM potential, optimized for liquid/solid interfaces, also accurately reproduces stacking fault energies and dislocation core geometries compared with density-functional theory calculations\cite{Yasi2009,Nogaret2010}.  The geometry of a $\bv = \aaxis$ kink of height $c$ is computed from a periodic cylinder with axis $N\aaxis + \caxis$; this produces a periodic array of kinks with density $N^{-1}$.  The inner cylinder radius is $22b$ plus an outer layer of $5b$ fixed to the initial positions from anisotropic elasticity for a mixed dislocation.  The initial geometry relaxes to a metastable mixed dislocation configuration.  To produce a kink, we (1) apply a small Escaig shear strain of $-0.006$ between the $\aaxis$ and $\baxis$ directions to promote basal spreading, (2) relax, (3) reverse the strain, and (4) relax.  The energy of the configurations varies as a linear function of $N$: constant kink energy plus the line energy with $N$.  The energies of the core cylinder at $N=60, 80, 100, 120$ go as $(0.515 + 0.910 N)$ eV, and we identify the single kink formation energy $\Hf = 0.515\eV$.  The core cylinder---where the total energy becomes linear in radius---is $12.5b$; if instead the sum is carried out to $22b$, the formation energy is 0.510eV.  The EAM potential has a zero-temperature prismatic Peierls stress $\taus = 140\MPa$; this is the minimum stress required for a straight $a$-type screw dislocation to glide in the prismatic plane.  We validate with the experimental data for cross-slip from 300--700K in \Sec{kink-geom}.

\TITLE{Results}
\SUBTITLE{Prismatic screw dislocation core}
\Fig{prismcore} shows the core of a prismatic screw dislocation from first-principles.  In Mg, the $a$-type screw dislocation splits into partials in the $(0001)$ plane; splitting onto the $(01\bar10)$ plane is a metastable higher energy configuration.  The basal stacking fault energy is 34 mJ/m$^2$ compared with the prismatic stacking fault energy of 218 mJ/m$^2$; the metastable point for the prismatic stacking fault is $\frac{a}{6}[11\bar20]+0.065c[0001]$.  We identify $\Sc=8$ sites in the core with large screw dislocation content where solutes may substitute and lower (or raise) the dislocation core energy.  Unlike the basal screw dislocation, there is very little volumetric strain and the dislocation core spreads rather than splitting into partials.  Because the prismatic core is metastable, solutes can destabilize the core; in that case, relaxation of the geometry will split the dislocation core on the basal plane.  Here we use direct calculation of interaction energy for attractive solutes.

\SUBTITLE{Direct solute/dislocation interaction}
\Fig{interaction} shows the difference in energy for solutes substituted at sites in a prismatic dislocation core relative to substitution in a basal dislocation core.  Three solutes decrease the prismatic stacking fault energy in Mg and produce attractive dislocation interactions: K, Na, and Sc.  Note that repulsive interactions are still possible as the sane site in the basal dislocation core can be \textit{more} attractive than in the prismatic dislocation core.  Of these three solutes, K has the strongest interaction in both dislocation cores due to the large size misfit and change in stacking fault energies (``chemical'' misfit\cite{Yasi2010}).  The statistics of the different site energies---mean site energy over the eight core sites $\DEave$ and standard deviation $\dE$---are included in \Tab{solute}.  These parameters enter into the analytic approximation to the statistical model we derive, and from which we compute the dimensionless softening parameter $\Pdk$ in \Sec{cross-slip}, $\taua/\taus$ in \Sec{migration}, and their ratio $\chi$.

\SUBTITLE{Kink geometry and enthalpy}
\label{sec:kink-geom}
\Fig{kink-geom} shows the relaxed kink geometry.  As kinks are ``decorated'' with solutes, we need to know the portion of the kink that is prismatic; these sites will have their energies changed by the energy difference between a prismatic and basal core containing a solute.  The kink has a height of $c$, and takes $30b$ to step from one basal plane to a neighboring basal plane; over that length, $15b$ is a prismatic core; hence, for $\lkink = 15$, we have $S = \Sc\lkink = 120$ sites which can be occupied by solute atoms.  We consider kinks with random distributions of solute, and derive the probability of having a kink with energy $E$; we then determine the distribution of double-kink nucleation energies and kink-migration barriers as well as the athermal stress required for kink mobility.

\Fig{cross-slip} compares the experimentally determined cross-slip stress for pure Mg\cite{WardFlynn1961} with the calculated double-kink nucleation enthalpy.  In a manner similar to Kocks~\et\cite{Kocks1975}, we expect the double-kink nucleation enthalpy to follow
\be
\Hdk(\tau) = 2\Hf\left\{1 - \left(\frac{\tau}{\taus}\right)^{1/2}\right\}
\label{eqn:Hdk}
\ee
where $\Hf$ is the formation energy of a single kink, and $\taus$ is the prismatic Peierls stress; the exponent of 1/2 corresponds to the elastic interaction of two well-separated kinks\cite{Pueschl2002}.  To validate this expression, we use the Orowan equation to relate the enthalpy to the plastic strain rate $\dot\eps$ for prismatic shear
\be
\dot\eps = \left(\frac{c}{a}\right)b^2\disl\nuf e^{-\Hdk(\tau)/\kB T}
\label{eqn:Orowan}
\ee
where $\disl$ is the dislocation density (taken as $10^8 \text{cm}^{-2}$), and $\nuf$ is the double kink nucleation attempt frequency.  The double-kink nucleation (or jog-pair\cite{Yoshinaga1963}) model for cross-slip is valid in the ``high'' (above room temperature) range\cite{Couret1985.1,Couret1985.2,Couret1991} where the stress is low enough that constriction on the prismatic plane is limited to the kinks.  At low temperatures, cross-slip occurs through bowing and this treatment does not apply.  The attempt frequency is difficult to compute accurately; however, it can be estimated as a typical phonon frequency ($\sim 10$THz) divided by $\sim 4S$ (the number of atoms in two kinks), or instead fit to the experimental data in the 300--700K range.  A single parameter fit with the experimental strain rate $\dot\eps=1.66\EE{-4}\text{s}^{-1}$ gives $\nuf = 15.4\text{GHz}$, which is remarkably close to our simple estimate counting degrees of freedom.  With this one parameter, we are able to reproduce the experimental cross-slip stress from room temperature up to $\sim400\C$,
\be
\tau(T) = \taus\left(1-\frac{\kB T}{\Hdkb}\right)^2
\label{eqn:pureMgcross-slip}
\ee
where
\be
\Hdkb = \frac{2\Hf}{\ln(ca\disl\nuf)-\ln\dot\eps} = 62.5\meV
\ee
for our dislocation density and strain rate.  For notational convenience, we introduce the parameter $\adk$ as
\be
\adk = 1 - \left(\frac{\tau}{\taus}\right)^{1/2};
\label{eqn:alpha}
\ee
for pure Mg, $\adk^0 = \kB T/\Hdkb$ from \Eqn{pureMgcross-slip}.  The excellent agreement with both tensile tests and in situ experiments validates our use of a double-kink nucleation model for cross-slip above room temperature; it should be noted that there is marked deviation going to absolute zero, as the mechanism for cross-slip changes.

\SUBTITLE{Distribution of double-kink nucleation energies}
\label{sec:cross-slip}
Each nucleated pair of kinks in a solute field requires a total energy equal to the formation energies plus the change due to the presence of solutes in the kink; the distribution can be computed numerically.  Each kink has $S=120$ sites in the prismatic core of the kink that may be occupied by solute atoms; we assume that the total energy change for the kink is the sum of all the individual energy changes, which are given by the site occupancy (either 0 or 1) multiplied by the energy of that site.  We assume translational invariance down the length of the kink, so that there are only $\Sc=8$ unique site energies to consider labeled $\DE_i$ for $i=1\ldots\Sc$---these are the site energy changes in \Fig{interaction}.  Then, in the kink core, each row of sites has $n_i$ solutes (between 0 and $\lkink=15$), which contribute energy $n_i\DE_i$; the total occupancy of the $S$ core sites is $n=\sum_i n_i$ and the energy is $E=\sum_i n_i\DE_i$.  Hence, we can write the number of possible configurations involving $n_0$ out of $S$ kink core sites occupied with energy change $E_0$ as
\be
g(E_0,n_0)=\sum_{n_1=0}^{\lkink}\cdots\sum_{n_{\Sc}=0}^{\lkink} \delta_{n,n_0}\delta(E_0-E) \prod_i \binom{\lkink}{n_i}
\label{eqn:DOS}
\ee
where $\delta(E_0-E)$ is the Dirac delta function, $\delta_{n,n_0}$ is the Kronecker delta, and the final term accounts for the multiplicity of occupancies along each row.  For numerical convenience, we approximate the delta function with a smoothed Gaussian with width 10meV.  From \Eqn{DOS}, the fraction of double kinks that can form with energy $2\Hf + E$ for a random solute distribution concentration $\conc$ is
\be
\Gdk(2\Hf + E,\conc) = \sum_{n=0}^S\sum_{m=0}^S \conc^{n+m}(1-\conc)^{2S-n-m} \int_{-\infty}^{\infty} dE' g(E',n) g(E-E',m).
\label{eqn:DOSdk}
\ee
\Fig{prob} shows the numerical double-kink nucleation energy distributions for several concentrations.  The average nucleation rate for double kinks at stress $\tau$ and temperature $T$ is
\be
\tdk^{-1} = \nuf \int_{-\infty}^{\infty} dE \Gdk(2\Hf + E,\conc)\exp\left[-\frac{2\Hf+E}{\kB T}\left\{1 - \left(\frac{\tau}{\taus}\right)^{1/2}\right\}\right]
\label{eqn:dk-rate}
\ee
assuming $\Gdk(2\Hf+E,\conc)\approx 0$ for $E<-2\Hf$; otherwise the integral must be split at $E=-2\Hf$.

An analytic approximation can be derived for the distribution of energies and the average nucleation time.  \Eqn{DOS} (and hence, \Eqn{DOSdk}) can be alternately viewed as the distribution of the sum of $n$ random variables, where each variable is the site energy.  From the central limit theorem, this distribution will be normal with mean value of $n\DEave$ and standard deviation $\sqrt{n}\dE$ (for average interaction $\DEave$ and standard deviation $\dE$; c.f.~\Tab{solute}),
\be
\Gdka(2\Hf + E,\conc) = \sum_{n=0}^{2S}\binom{2S}{n}\conc^{n}(1-\conc)^{2S-n} \frac{1}{\sqrt{2\pi n}\dE}\exp\left[-\frac{(E-n\DEave)^2}{2n\dE^2}\right]
\label{eqn:DOSdka}
\ee
This form matches the appearance of distributions in \Fig{prob}.  The analytic expression of \Eqn{DOSdka} can integrated in \Eqn{dk-rate}, using $\beta=(\kB T)^{-1}$ and $\adk$ from \Eqn{alpha}
\be
\begin{split}
\tdk^{-1} &\approx \nuf\sum_{n=0}^{2S}\binom{2S}{n}\conc^{n}(1-\conc)^{2S-n} \int_{-\infty}^{\infty} dE \frac{\exp\left[-\beta\adk(2\Hf+E) -\frac{(E-n\DEave)^2}{2n\dE^2}\right]}{\sqrt{2\pi n}\dE}\\
&= \nuf e^{-\beta\adk2\Hf}\sum_{n=0}^{2S}\binom{2S}{n}\conc^{n}(1-\conc)^{2S-n} \left[e^{-\beta\adk\DEave}e^{(\beta\adk\dE)^2/2}\right]^n\\
&= \nuf e^{-\beta\adk2\Hf}\left\{1-\conc+\conc e^{-\beta\adk\DEave}e^{(\beta\adk\dE)^2/2}\right\}^{2S}
\end{split}
\ee
where the first term is the nucleation rate in the absence of solutes, and the term in braces is the change in rate due to solutes.

The analytic rate equation can be simplified to give the softening stress with concentration in the low concentration limit.  The change in cross-slip stress can be understood as a change in the stress necessary to have the \textit{same} nucleation rate at a given temperature in the \textit{absence} of solutes; that is, the cross-slip stress $\tau$ has a corresponding $\adk$ such that
\be
e^{-\beta\adk^0 2\Hf} = e^{-\beta\adk 2\Hf}\left\{1-\conc+\conc e^{-\beta\adk\DEave}e^{(\beta\adk\dE)^2/2}\right\}^{2S}
\ee
where $\adk^0=\kB T/\Hdkb$.  For small $\conc$, $d\adk/d\conc$ is
\be
\begin{split}
e^{\beta(\adk-\adk^0)2\Hf/(2S)} &= 1-\conc+\conc e^{-\beta\adk\DEave}e^{(\beta\adk\dE)^2/2}\\
1+\frac{\beta 2\Hf}{2S}(\adk-\adk^0) &\approx 1-\conc+\conc e^{-\beta\adk^0\DEave}e^{(\beta\adk^0\dE)^2/2}\\
\frac{d\adk}{d\conc} &\approx \frac{2S}{\beta2\Hf}\left[
\exp\left(-\frac{\DEave}{\Hdkb} + \frac12\left(\frac{\dE}{\Hdkb}\right)^2\right)-1\right]
\end{split}
\ee
Finally, as $\tau = \taus(1-\adk)^2$, the change in cross-slip stress is
\be
\begin{split}
\frac{d\tau}{d\conc} &= -\taus 2 (1-\adk)\frac{d\adk}{d\conc}\\
&= -\taus (1-\adk^0)\adk^0\cdot\left(\frac{\Hdkb}{2\Hf}4S\right)\left[
\exp\left(-\frac{\DEave}{\Hdkb} + \frac12\left(\frac{\dE}{\Hdkb}\right)^2\right)-1\right]\\
&\equiv -\taus (1-\adk^0)\adk^0\cdot \Pdk
\end{split}
\label{eqn:Pdk}
\ee
where $\Pdk$ is the unitless ``softening potency'' (c.f.~\Tab{solute}).  Note that  $\Pdk>0$ requires $\DEave < \frac12 \Hdkb (\dE/\Hdkb)^2$, so that solutes with repulsive interactions \textit{can} soften cross-slip provided sufficient attractive sites are available.

\SUBTITLE{Distribution of kink migration energy barriers}
\label{sec:migration}
Solutes ``roughen'' the energy landscape for kinks and provide two barriers to the motion of kinks: a minimum (athermal) stress required for kinks to migrate preferentially down the dislocation line, and the energy barrier over the length of a kink\cite{Argon2005}.  Solutes provide local changes in energy as a kink moves over a single Burgers vector; a minimum stress---$\tauac(\conc)$---is necessary to overcome this short-range change in energy.  This is given by the average roughness of the energy landscape: the energy changes as $\Sc$ sites with solutes ``leave'' the kink and another $\Sc$ sites ``enter'' the kink; the energy change for solutes entering is $+\DE_i$, and is $-\DE_i$ for those leaving.  If the width of the solute interaction is approximated as a Gaussian with area $4b^2$, the stress to overcome should go as $\sqrt{2/e} \cdot(\text{energy difference})/4b^3$.  Analytically, we consider a distribution of energy differences as the sum of $n$ random variables for the energies; however, both $\pm\DE_i$ are equally likely so the mean is 0 and standard deviation $\dDE^{1/2}$.  The average energy difference is symmetric around $n=\Sc$; hence, the standard deviation of our normal distribution for $n=0..\Sc$ is $\sqrt{n}\dDE^{1/2}$ and for $n=\Sc..2\Sc$ it is $\sqrt{2\Sc-n}\dDE^{1/2}$.  From this, we can approximate the average absolute energy change $\DEabs$ as the kink moves by one lattice spacing,
\be
\begin{split}
\DEabs =& \Bigg[\sum_{n=0}^{\Sc} \binom{2\Sc}{n} \conc^n (1-\conc)^{2\Sc-n} \sqrt{n}\\
&+ \sum_{n=\Sc+1}^{2\Sc} \binom{2\Sc}{n} \conc^n(1-\conc)^{2\Sc-n} \sqrt{2S-n}\Bigg]\sqrt{\frac{2}{\pi}} \dDE^{1/2}\\
\approx& \sqrt{\frac{2}{\pi}}(2\Sc)\conc\dDE^{1/2}
\end{split}
\label{eqn:Hath}
\ee
and then the athermal stress is
\be
\tauac(\conc) = \sqrt{\frac{2}{e}}\frac{\DEabs}{4b^3}
\approx \frac{2}{\sqrt{e\pi}}\frac{2\Sc}{4b^3}\dDE^{1/2}\conc \equiv \taua \conc
\label{eqn:athermal}
\ee
where the last expression is a simple analytic model in the limit of low concentration.  The numerical distribution of energies is the athermal distribution in \Fig{prob}.  In addition to this athermal stress, there are energy barriers as a kink moves by its length $\lkink$ due to solute occupancy changes. For a random distribution of solutes at concentration $\conc$, the fraction of energy barriers $E$ is
\be
\Gkm(E,\conc) = \sum_{n=0}^S\sum_{m=0}^S \conc^{n+m}(1-\conc)^{2S-n-m} \int_{-\infty}^{\infty} dE' g(E',n) g(E+E',m).
\label{eqn:DOSkm}
\ee
Note the sign change for $g(E+E',m)$ compared with \Eqn{DOSdk}.  The enthalpy barrier to escape an energy well $E>0$ is (assuming a Gaussian of width $2\lkink b$ and kink height $h=c$)
\be
E\left[1-\frac{2\lkink b^2h}{\sqrt{2/e}}\frac{\tau-\tauac(\conc)}{E}\right]^{3/2}
\label{eqn:kmenthalpy}
\ee
for $\tau>\tauac(\conc)$, from \cite{Argon2005}.  Define
\be
\Htau = (\tau-\tauac(\conc))\frac{2\lkink b^2}{\sqrt{2/e}}
\ee
then the time needed to overcome all barriers along the dislocation line is
\be
\tkm = \nuf^{-1}\left\{\int_{-\infty}^{\Htau} dE \Gkm(E,\conc) + \int_{\Htau}^{\infty} dE \Gkm(E,\conc)\exp\left(\frac{E}{\kB T}\left[1-\frac{\Htau}{E}\right]^{3/2}\right)\right\}
\label{eqn:tkm}
\ee
Note that $\tkm$ is finite only when $\tau>\tauac(\conc)$.

There is a minimum solute concentration---and hence, athermal stress---necessary for thermally-activated kink migration to affect the cross-slip stress.  \Eqn{DOSkm} is a distribution of the sum of $n$ random energies; however, as solutes enter and leave the kink, our possible energies are $\pm\DE_i$ for a distribution with mean 0 and standard deviation $\dDE^{1/2}$ as with the athermal barrier.  Due to symmetry around $n=S$, the standard deviation of our normal distribution for $n_0=0..S$ is $\sqrt{n}\dDE^{1/2}$ and for $n=S..2S$ it is $\sqrt{2S-n}\dDE^{1/2}$.  From this, we approximate $\Gkm(E,\conc)$ as normal with mean 0 and variance
\be
\begin{split}
\Hkm =& \Bigg[\sum_{n=0}^S \binom{2S}{n} \conc^n (1-\conc)^{2S-n} n
+ \sum_{n=S+1}^{2S} \binom{2S}{n} \conc^n(1-\conc)^{2S-n} (2S-n)\Bigg]\dDE\\
\approx& (2S)\conc\dDE
\end{split}
\label{eqn:Hkm}
\ee
where the approximation has less than 1\%\ error for $\conc\lesssim45\%$ with $S=120$.  Hence, the standard deviation is $\approx \sqrt{2S}\dDE^{1/2}\conc^{1/2}$.  With this approximation, the stress necessary for thermally activated kink migration to require a larger stress than athermal kink migration is when $\tkm^{-1}$ at $\tau=\tauac(\conc)$ is slower than required by the Orowan equation: $\dot\eps> ca\disl \tkm^{-1}$ (c.f.~\Eqn{Orowan}).  At the athermal stress, there is no reduction in enthalpy for $E>0$, and so
\be
\begin{split}
\tkm\nuf &= \frac12 + \int_0^{\infty} dE \Gkm(E,\conc)e^{\beta E}\\
&\approx \frac12 + \int_0^{\infty} dE \frac{\exp\left\{\beta E - E^2/\left[2\Hkm \right]\right\}}{\sqrt{2\pi\Hkm}}\\
&\approx \exp\left[\frac12 \beta^2 (2S)\dDE\conc\right]
\end{split}
\ee
where the first approximation is the use of a normal distribution and the second is valid when the exponential term is larger than 1.  Then, the minimum concentration $\concmin$ is such that $\dot\eps=ca\disl\tkm^{-1}$; above this concentration, thermally activated kink migration will be required.  Then
\be
\begin{split}
\frac12 \beta^2 (2S)\dDE\concmin &= \ln\left(\frac{ca\disl\nuf}{\dot\eps}\right)\\
\concmin &= \frac{(\kB T)^2}{\dDE} \frac{2\Hf}{\Hdkb S}.
\end{split}
\ee
We rewrite this in terms of the minimum value of \text{athermal} stress, as $\tauac(\conc)\approx \taua \conc$,
\be
\tauacmin \approx \taua\concmin
=\frac{(\kB T)^2\dDE^{-1/2}}{\sqrt{e\pi}b^3\lkink}\frac{2\Hf}{\Hdkb}.
\label{eqn:crossover}
\ee
If $\tauac(\conc)<\tauacmin$, then thermally activated kink-migration will lead to further strengthening.  For the solutes we consider here (c.f.~\Tab{solute} and \Fig{solute-cross-slip}), $\tauacmin$ is $\gtrsim 15\MPa$ at 300K and $\gtrsim 40\MPa$ at 600K; hence, thermally-activated kink-migration only limits softening outside of the stress and temperature range of interest.  Above $\concmin$, thermally-activated kink-migration controls cross-slip.  The integral in \Eqn{tkm} can be computed in closed form with the approximation
\beu
\Gkm(E,\conc)\approx\exp(-E^2/(2\Hkm))/\sqrt{2\pi\Hkm}
\eeu
from \Eqn{Hkm} and with
\be
\left(1-\frac{\Htau}{E}\right)^{3/2}\approx
\begin{cases}
(2-\sqrt{3})\left(\frac{E-\Htau}{\Htau}\right)&: \Htau<E<(3+\sqrt{3})\Htau/2\\
1-\frac32\frac{\Htau}{E}&: E>(3+\sqrt{3})\Htau/2
\end{cases}
\ee
The closed-form expression for \Eqn{tkm} is unwieldy, and omitted for clarity.

\SUBTITLE{Prediction of cross-slip stress with concentration and temperature}
\Fig{solute-cross-slip} shows the predicted cross-slip stress with concentration and temperature predicted by our numerical solution of the Orowan equation, and analytic approximations for softening and athermal hardening.  The Orowan equation, \Eqn{Orowan}, is modified for the two thermally-activated cross-slip processes as
\be
\dot\eps = \left(\frac{c}{a}\right)b^2\disl\left(\tdk^{-1}+\tkm^{-1}\right)^{-1}.
\label{eqn:Orowannew}
\ee
This relates the plastic strain rate to the time to nucleate a pair of double-kinks and migrate the length of the dislocation line.  The analytic curves are for double-kink nucleation and athermal hardening, both of which are approximately linear.  For all solutes, the analytic approximations are reasonable at lower $\conc$, with more significant deviations at higher concentrations.  All three solutes show softening for low concentrations which becomes hardening at higher concentrations; the range of solute concentration that leads to softening decreases at higher temperatures.  These predictions suggest possible alloying concentrations that can lead to lower stress for thermally-activated cross-slip, and hence decrease the plastic anisotropy.  At 600K, the cross-slip stress is 4.1MPa from our model; that cross-slip stress occurs at 545K for Mg-0.4at.\%K, 560K for Mg-0.6at.\%Na, and 565K for Mg-0.7at.\%Sc.

We derive approximate analytic expressions for the dilute concentration limit for alloy design. The optimal solute concentration $\concopt$ occurs when the minimum stress for double-kink nucleation matches the athermal stress; hence,
\be
\begin{split}
\taus(1-\adk^0)(1-\adk^0-\adk^0\Pdk\concopt) &= \taua\concopt \\
(1-\adk^0)^2 - (1-\adk^0)\adk^0\Pdk\concopt &= \frac{\taua}{\taus}\concopt\\
\frac{(1-\adk^0)^2}{(1-\adk^0)\adk^0\Pdk + \taua/\taus} &= \concopt
\end{split}
\label{eqn:concopt}
\ee
and then the cross-slip stress is $\taua\concopt$.  Finally, the cross-slip stress for pure Mg at one temperature can now be achieved at a minimum ``equivalent temperature'' in an alloy.  The pure Mg forming temperature $T_{\text{form}}$ is written as $\adk^0=\kB T_{\text{form}}/\Hdkb$, and the minimum equivalent temperature $T_{\text{minimum}}$ as $\adk=\kB T_{\text{minimum}}/\Hdkb$; then,
\be
\begin{split}
\taus(1-\adk^0)^2 &= \frac{(1-\adk)^2\taua}{(1-\adk)\adk\Pdk + \taua/\taus}\\
\left[\left(\frac{1-\adk}{1-\adk^0}\right)^2-1\right] &= \frac{\Pdk}{\taua/\taus}(1-\adk)\adk.
\end{split}
\label{eqn:equivtemp}
\ee
Define $\chi=\Pdk/(\taua/\taus)$, the unitless ratio of softening potency to athermal hardening; the higher this parameter, the more the forming temperature of the alloy can be lowered.  The solution to the quadratic formula is
\be
1-\adk = \frac{\frac12\chi(1-\adk^0)^2 + \sqrt{\frac14(\chi(1-\adk^0)^2)^2 + (1-\adk^0)^2 + \chi(1-\adk^0)^4}}{1+\chi(1-\adk^0)^2}
\label{eqn:equivtemp-quad}
\ee
This is plotted in \Fig{equivtemp} for a variety of $\chi$ values.  Comparing to our numerical values, the analytic approximation gives equivalent 600K forming temperatures  of 530K for Mg-K, 543K for Mg-Na, and 560K for Mg-Sc.  Note that for $T_{\text{form}}=600\text{K}$ and $T_{\text{minimum}}=300\text{K}$ (room temperature forming) requires $\chi=43.3$, which would be an attractive interaction to a kink of at least $\sim 4\Hdkb = 250\meV$.

\TITLE{Conclusions}
Our first-principles calculations of solute interaction with prismatic and basal dislocations combined with our computational model of basal to prismatic cross-slip above room temperature allows us to make the first prediction of cross-slip softening of Mg in three binary alloys with lower forming temperatures.  This model accounts for the random distribution of solutes in a dislocation core with different possible energies, and the direct calculations can be used above the dilute concentration limit.  At the same time, an approximate analytic model connects average solute interaction energies to the dilute limit for near-quantitative predictions of cross-slip softening, and to determine the required interaction energies that lead to softening.  This predicts possible formable Mg alloys and sets the stage for future work covering a wider range of solutes; moreover, the aspects of the approach can be applied to other thermally-activated plastic deformation mechanisms.  Replacing direct calculations of interaction energies with a geometric model connecting the prismatic dislocation core to changes in prismatic stacking fault energy could also treat other substitutional solutes.

\subsection*{Acknowledgments}
This research was sponsored by NSF through the GOALI program, grant 0825961, and with the support of General Motors, LLC, and in part by NSF through TeraGrid resources provided by NCSA and TACC, and with a donation from Intel.  \Fig{kink-geom} was rendered with VMD\protect\cite{Humphrey1996}.

\newpage

\begin{table}[p]
\caption{Direct solute/kink interaction energy statistics and parameters for analytic approximations.  The analytic softening and hardening depend on the distribution of solute energies around a kink; from these, the (unitless) solute softening potency $\Pdk$ and (unitless) athermal slip prefactor $\taua/\taus$ are derived.  The ratio of the two factors, $\chi$, determines the maximal amount of softening that is possible in a given alloy; larger $\chi$ values indicates more potential softening (c.f.~\Fig{equivtemp}).  The linear analytic approximations are reasonable for $\conc\lesssim 2\%$ (c.f.~\Fig{solute-cross-slip}).}
\label{tab:solute}
\begin{center}
\begin{tabular}{lccccccc}
&\multicolumn{3}{c}{interaction [meV]}&\multicolumn{3}{c}{strength param.}\\
& $\DEave$ & $\dE$ & $\dDE^{1/2}$ & $\Pdk$ & $\taua/\taus$ & $\chi$\\
\hline
K & --45.8 & 44.0 & 63.5 & 48.5 & 6.67 & 7.27 \\
Na& --37.0 & 22.3 & 43.2 & 26.9 & 4.54 & 5.93 \\
Sc& --19.6 & 38.1 & 42.8 & 18.9 & 4.50 & 4.20 
\end{tabular}
\end{center}
\end{table}

\bfig
\centering
\includegraphics[width=\smallfigwidth]{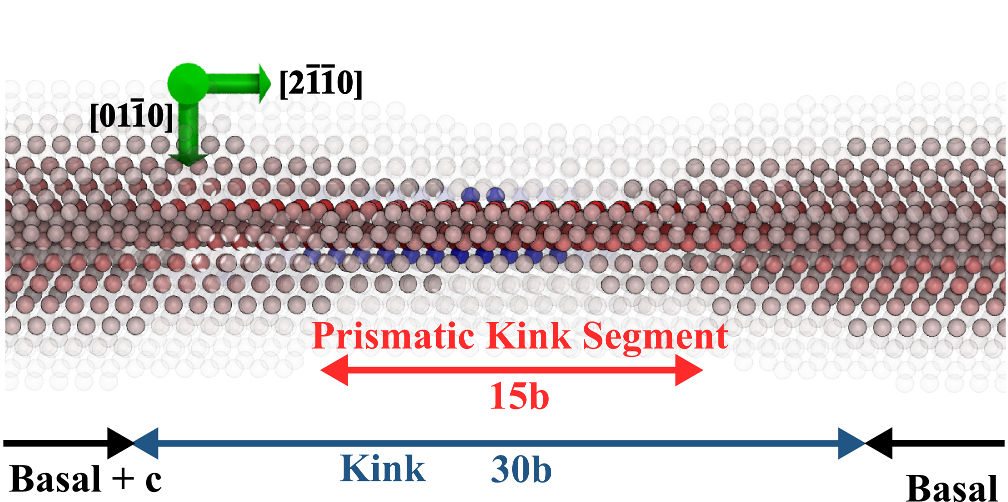}
\includegraphics[width=\smallfigwidth]{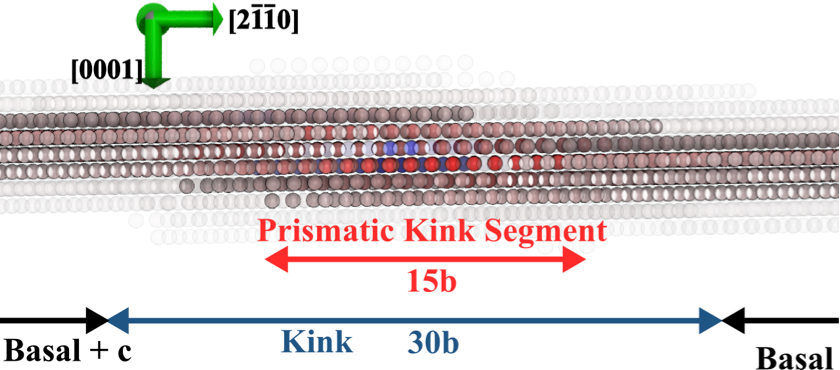}
\caption{Kink geometry from Sun potential.  The atom colors indicates changes in atomic energy relative to bulk Mg.}
\label{fig:kink-geom}
\efig

\bfig
\centering
\includegraphics[width=\smallfigwidth]{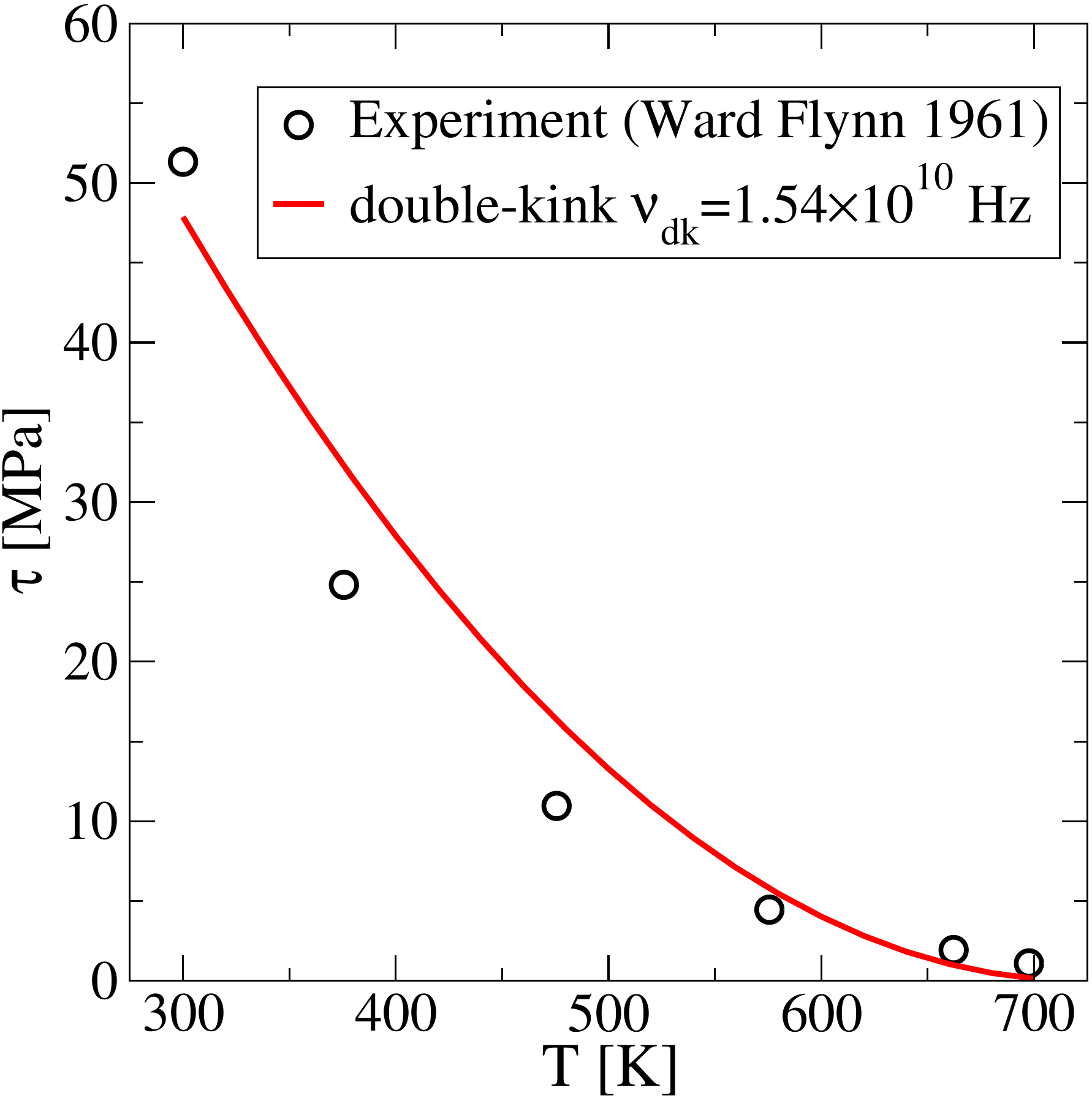}
\caption{Cross-slip stress with temperature from experiment and double-kink nucleation enthalpy value of \Eqn{pureMgcross-slip}.  The single kink formation energy and prismatic Peierls stress are computed directly from the Mg EAM potential; the Orowan equation determines a cross-slip stress for a given dislocation density and plastic strain rate.  There is a single unknown parameter $\nuf$ which is determined by a least-squares fit to the experimental data; simple counting of degrees of freedom suggests an order of magnitude of $\sim 10^{10}\text{Hz}$ as found in our fit.}
\label{fig:cross-slip}
\efig

\bfig
\centering
\includegraphics[width=2.5in]{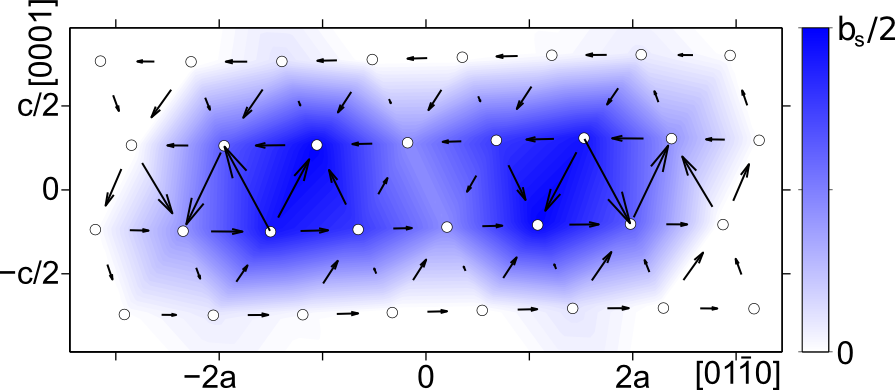}
\includegraphics[height=2in]{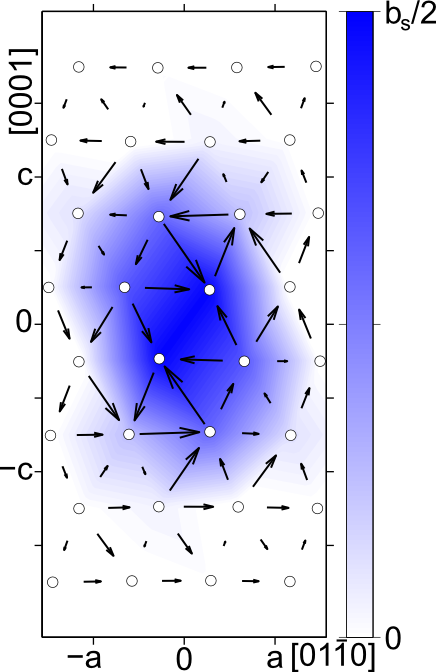}
\caption{Basal and prismatic $a$-type screw dislocation core geometries.  The differential displacement plots (arrows) indicate relative displacement of rows by up to $b/4$ to show the small amount of splitting into partials.  The coloring is a linear-interpolation of Nye tensor density indicating up to a total displacement of $b/2$.  The metastable prismatic core has a small amount of spreading in the prismatic plane.}
\label{fig:prismcore}
\efig

\bfigwide
\centering
\includegraphics[width=0.3\wholefigwidth]{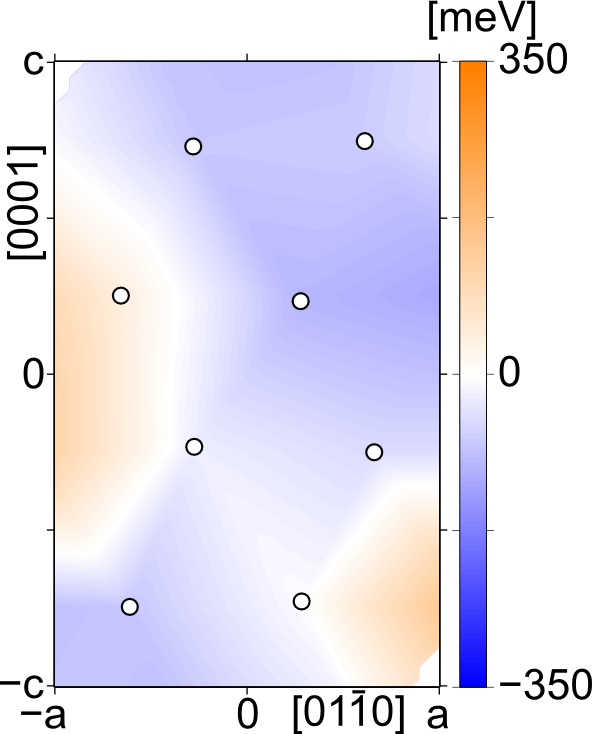}
\includegraphics[width=0.3\wholefigwidth]{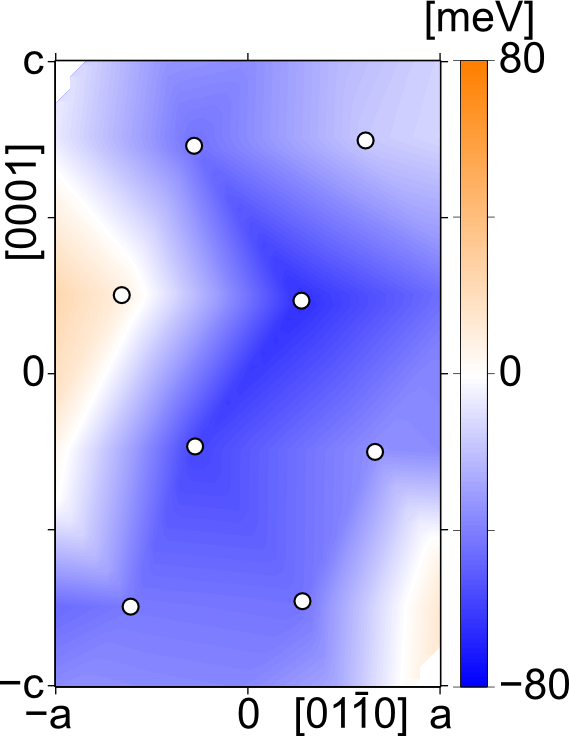}
\includegraphics[width=0.3\wholefigwidth]{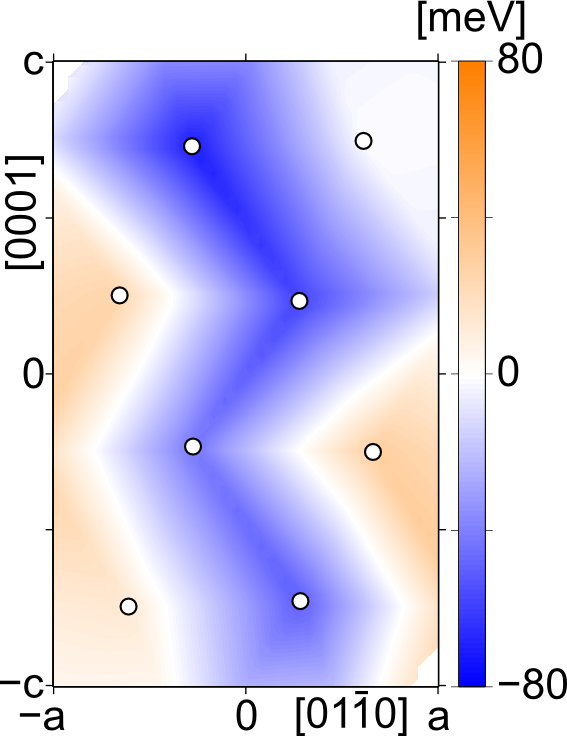}
\caption{The relative energy of a K, Na, and Sc solute in a prismatic dislocation core compared with the same site in a basal dislocation core.  The change in energy can promote (negative energies) or prevent (positive energies) the formation of a double kink.  The interaction energy of a single solute with a single dislocation enters into the distribution of double-kink formation energies, kink-migration barrier energies, and athermal kink migration stress.}
\label{fig:interaction}
\efigwide

\bfigwide
\centering
\includegraphics[width=\wholefigwidth]{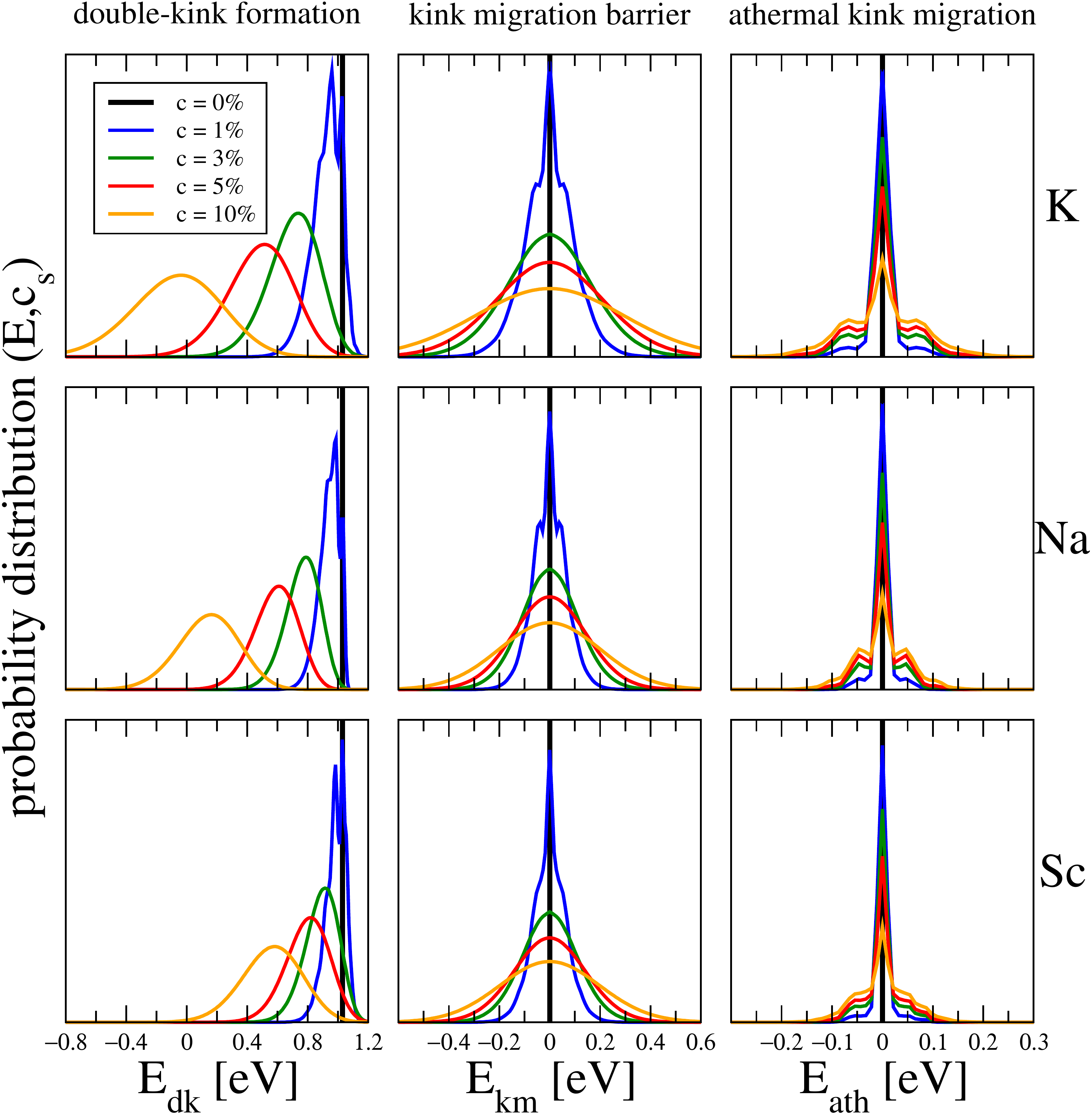}
\caption{Probability distributions for all three processes (double-kink nucleation, thermally-activated kink migration, and athermal kink migration) and solutes (K, Na, Sc).  The probability distributions are determined numerically from the solute interaction energies assuming random solute distributions.  For ``large'' solute $\conc$ (compared to the inverse of the number of available kink size $S^{-1}$), the distributions become close to normal, and approximate analytic expressions can be derived.}
\label{fig:prob}
\efigwide

\bfigwide
\centering
\includegraphics[width=\wholefigwidth]{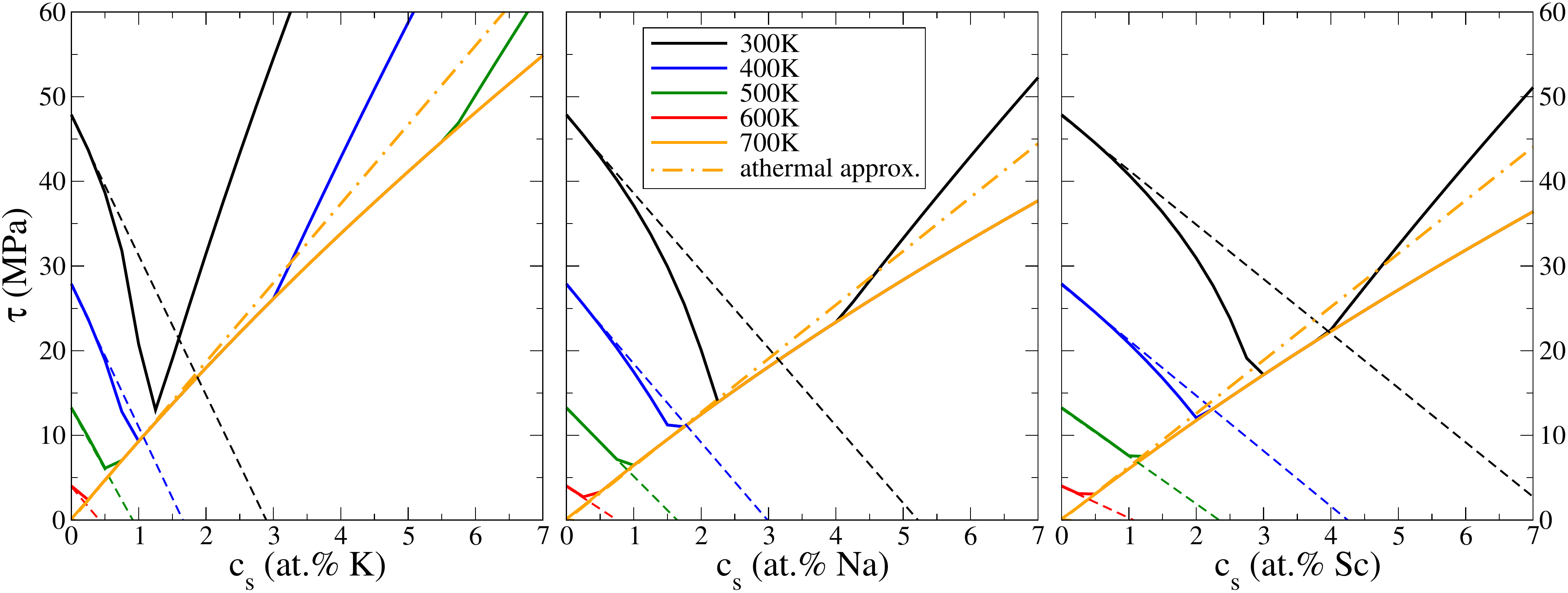}
\caption{Cross-slip stress with concentration for K, Na, and Sc between 300--700K.  Increasing temperature lowers the pure Mg cross-slip stress; small additions of attractive solutes increase the double-kink nucleation rate, leading to softening.  The lower bound for all curves is the athermal stress $\tauac(\conc)$; only at higher concentrations does thermally-activated kink-migration need to be activated to achieve the required plastic strain rate.  The dashed lines show the linear softening analytic approximation, and the dashed-dotted curves are the linear analytic approximation to the athermal stress.  All cases show a minimum cross-slip stress for a given temperature, and the concentration range for cross-slip softening decreases with increasing concentration.  This allows us to identify an ``equivalent temperature'': the lowest temperature where the binary alloy has the same cross-slip stress as pure Mg has at a higher temperature.  Hence, forming of these binary alloys can be done at a lower temperature, provided the solute concentration is properly tuned.}
\label{fig:solute-cross-slip}
\efigwide

\bfig
\centering
\includegraphics[width=\smallfigwidth]{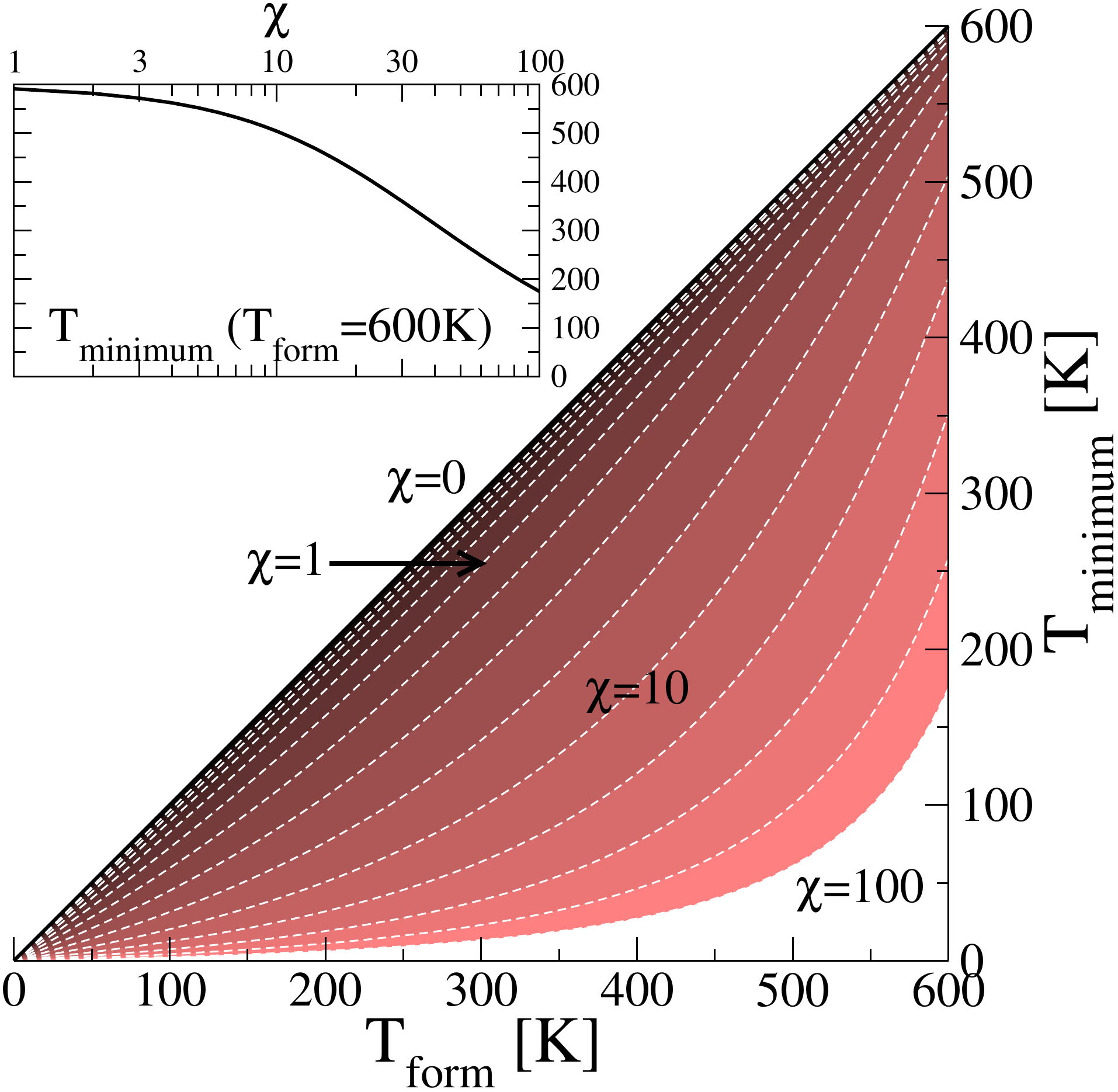}
\caption{Analytic equivalent minimum temperature for different solute interaction parameters $\chi=\Pdk/(\taua/\taus)$.  Contours for each $\chi$ are shown, increasing logarithmically.  In the analytic approximation, the ratio of double-kink softening to athermal hardening determines how effective a solute can be at reducing the cross-slip stress for a Mg alloy.  The forming temperature for pure Mg defines the necessary cross-slip stress; the minimum temperature that a solute can give the \textit{same} cross-slip stress is plotted against that forming temperature for a range of parameters.  The inset shows the decrease in forming temperature relative to 600K.  Note: the thermally-activated cross-slip model is only valid for temperatures above room temperature, and the analytic approximations are accurate for $\conc\lesssim 2\%.$}
\label{fig:equivtemp}
\efig


\begin{thebibliography}{33}
\providecommand{\natexlab}[1]{#1}
\providecommand{\url}[1]{\texttt{#1}}
\providecommand{\urlprefix}{URL }
\expandafter\ifx\csname urlstyle\endcsname\relax
  \providecommand{\doi}[1]{doi:\discretionary{}{}{}#1}\else
  \providecommand{\doi}{doi:\discretionary{}{}{}\begingroup
  \urlstyle{rm}\Url}\fi
\providecommand{\eprint}[2][]{\url{#2}}
\providecommand{\BIBand}{and}
\providecommand{\bibinfo}[2]{#2}
\ifx\xfnm\undefined \def\xfnm[#1]{\unskip,\space#1}\fi
\bibitem[{Friedrich and Mordike(2006)}]{MgTech2006}
\bibinfo{author}{Friedrich\xfnm[ H.E.]}, \bibinfo{author}{Mordike\xfnm[ B.L.]}.
\newblock \bibinfo{title}{Magnesium Technology: Metallurgy, Design Data,
  Applications}.
\newblock \bibinfo{address}{Berlin Heidelberg}:
  \bibinfo{publisher}{Springer-Verlag}; \bibinfo{year}{2006}.
\bibitem[{Pollock(2010)}]{Pollock2010}
\bibinfo{author}{Pollock\xfnm[ T.M.]}.
\newblock \bibinfo{journal}{Science}
  \bibinfo{year}{2010};\bibinfo{volume}{328}(\bibinfo{number}{5981}):\bibinfo{%
pages}{986--987}.
\newblock \doi{\bibinfo{doi}{10.1126/science.1182848}}.
\bibitem[{Taylor(1938)}]{Taylor1938}
\bibinfo{author}{Taylor\xfnm[ G.I.]}.
\newblock \bibinfo{journal}{J Inst Metals}
  \bibinfo{year}{1938};\bibinfo{volume}{62}:\bibinfo{pages}{307--338}.
\bibitem[{Agnew and Duygulua(2005)}]{Agnew2005}
\bibinfo{author}{Agnew\xfnm[ S.R.]}, \bibinfo{author}{Duygulua\xfnm[ {\"O}.]}.
\newblock \bibinfo{journal}{Int J Plast}
  \bibinfo{year}{2005};\bibinfo{volume}{21}:\bibinfo{pages}{1161--1193}.
\newblock \doi{\bibinfo{doi}{10.1016/j.ijplas.2004.05.018}}.
\bibitem[{Akhtar and Teghtsoonian(1969)}]{Akhtar1969b}
\bibinfo{author}{Akhtar\xfnm[ A.]}, \bibinfo{author}{Teghtsoonian\xfnm[ E.]}.
\newblock \bibinfo{journal}{Acta Metall}
  \bibinfo{year}{1969};\bibinfo{volume}{17}:\bibinfo{pages}{1351--1356}.
\bibitem[{Ahmadieh et~al.(1965)Ahmadieh, Mitchell and Dorn}]{Ahmadieh1965}
\bibinfo{author}{Ahmadieh\xfnm[ A.]}, \bibinfo{author}{Mitchell\xfnm[ J.]},
  \bibinfo{author}{Dorn\xfnm[ J.E.]}.
\newblock \bibinfo{journal}{Trans Met Soc AIME}
  \bibinfo{year}{1965};\bibinfo{volume}{233}:\bibinfo{pages}{1130--1137}.
\bibitem[{Urakami et~al.(1970{\natexlab{a}})Urakami, Meshii and
  Fine}]{Urakami1970a}
\bibinfo{author}{Urakami\xfnm[ A.]}, \bibinfo{author}{Meshii\xfnm[ M.]},
  \bibinfo{author}{Fine\xfnm[ M.E.]}.
\newblock \bibinfo{journal}{Acta metall}
  \bibinfo{year}{1970}{\natexlab{a}};\bibinfo{volume}{18}:\bibinfo{pages}{87--%
99}.
\bibitem[{Urakami et~al.(1970{\natexlab{b}})Urakami, Meshii and
  Fine}]{Urakami1970b}
\bibinfo{author}{Urakami\xfnm[ A.]}, \bibinfo{author}{Meshii\xfnm[ M.]},
  \bibinfo{author}{Fine\xfnm[ M.E.]}.
\newblock \bibinfo{journal}{Proc 2nd ICMSA}
  \bibinfo{year}{1970}{\natexlab{b}};\bibinfo{volume}{1}:\bibinfo{pages}{272--%
276}.
\bibitem[{Urakami and Fine(1971)}]{Urakami1971}
\bibinfo{author}{Urakami\xfnm[ A.]}, \bibinfo{author}{Fine\xfnm[ M.E.]}.
\newblock \bibinfo{journal}{Acta metall}
  \bibinfo{year}{1971};\bibinfo{volume}{19}:\bibinfo{pages}{887--894}.
\bibitem[{Pink and Arsenault(1979)}]{Pink1979}
\bibinfo{author}{Pink\xfnm[ E.]}, \bibinfo{author}{Arsenault\xfnm[ R.J.]}.
\newblock \bibinfo{journal}{Prog Mater Sci}
  \bibinfo{year}{1979};\bibinfo{volume}{24}:\bibinfo{pages}{1--50}.
\bibitem[{Couret and Caillard(1985{\natexlab{a}})}]{Couret1985.1}
\bibinfo{author}{Couret\xfnm[ A.]}, \bibinfo{author}{Caillard\xfnm[ D.]}.
\newblock \bibinfo{journal}{Acta metall}
  \bibinfo{year}{1985}{\natexlab{a}};\bibinfo{volume}{33}:\bibinfo{pages}{1447%
--1454}.
\bibitem[{Couret and Caillard(1985{\natexlab{b}})}]{Couret1985.2}
\bibinfo{author}{Couret\xfnm[ A.]}, \bibinfo{author}{Caillard\xfnm[ D.]}.
\newblock \bibinfo{journal}{Acta metall}
  \bibinfo{year}{1985}{\natexlab{b}};\bibinfo{volume}{33}:\bibinfo{pages}{1455%
--1461}.
\bibitem[{Yoshinaga and Horiuchi(1963)}]{Yoshinaga1963}
\bibinfo{author}{Yoshinaga\xfnm[ H.]}, \bibinfo{author}{Horiuchi\xfnm[ R.]}.
\newblock \bibinfo{journal}{Trans Japan Inst Metals}
  \bibinfo{year}{1963};\bibinfo{volume}{5}:\bibinfo{pages}{14}.
\bibitem[{Couret et~al.(1991)Couret, Caillard, P{\"u}schl and
  Schoeck}]{Couret1991}
\bibinfo{author}{Couret\xfnm[ A.]}, \bibinfo{author}{Caillard\xfnm[ D.]},
  \bibinfo{author}{P{\"u}schl\xfnm[ W.]}, \bibinfo{author}{Schoeck\xfnm[ G.]}.
\newblock \bibinfo{journal}{Philos Mag A}
  \bibinfo{year}{1991};\bibinfo{volume}{63}:\bibinfo{pages}{1045--1057}.
\bibitem[{Friedel(1959)}]{Friedel1959}
\bibinfo{author}{Friedel\xfnm[ J.]}.
\newblock In: \bibinfo{editor}{Rassweiler\xfnm[ G.M.]},
  \bibinfo{editor}{Grube\xfnm[ W.L.]}, editors. \bibinfo{booktitle}{Internal
  Stresses and Fatigue in Metals}. \bibinfo{publisher}{Elsevier};
  \bibinfo{year}{1959}, p. \bibinfo{pages}{220}.
\bibitem[{Escaig(1968)}]{Escaig1968}
\bibinfo{author}{Escaig\xfnm[ B.]}.
\newblock \bibinfo{journal}{Phys Stat Sol}
  \bibinfo{year}{1968};\bibinfo{volume}{28}:\bibinfo{pages}{463--474}.
\bibitem[{Sun et~al.(2006)Sun, Mendelev, Becker, Kudin, Haxhimali, Asta
  et~al.}]{Sun2006}
\bibinfo{author}{Sun\xfnm[ D.Y.]}, \bibinfo{author}{Mendelev\xfnm[ M.I.]},
  \bibinfo{author}{Becker\xfnm[ C.A.]}, \bibinfo{author}{Kudin\xfnm[ K.]},
  \bibinfo{author}{Haxhimali\xfnm[ T.]}, \bibinfo{author}{Asta\xfnm[ M.]},
  et~al.
\newblock \bibinfo{journal}{Phys Rev B}
  \bibinfo{year}{2006};\bibinfo{volume}{73}:\bibinfo{pages}{024116}.
\newblock \doi{\bibinfo{doi}{10.1103/PhysRevB.73.024116}}.
\bibitem[{Kresse and Hafner(1993)}]{Kresse93}
\bibinfo{author}{Kresse\xfnm[ G.]}, \bibinfo{author}{Hafner\xfnm[ J.]}.
\newblock \bibinfo{journal}{Phys Rev B}
  \bibinfo{year}{1993};\bibinfo{volume}{47}(\bibinfo{number}{1}):\bibinfo{page%
s}{RC558--561}.
\bibitem[{Kresse and Furthm{\"u}ller(1996)}]{Kresse96b}
\bibinfo{author}{Kresse\xfnm[ G.]}, \bibinfo{author}{Furthm{\"u}ller\xfnm[
  J.]}.
\newblock \bibinfo{journal}{Phys Rev B}
  \bibinfo{year}{1996};\bibinfo{volume}{54}(\bibinfo{number}{16}):\bibinfo{pag%
es}{11169--11186}.
\bibitem[{Bl{\"o}chl(1994)}]{Blochl1994}
\bibinfo{author}{Bl{\"o}chl\xfnm[ P.E.]}.
\newblock \bibinfo{journal}{Phys Rev B}
  \bibinfo{year}{1994};\bibinfo{volume}{50}:\bibinfo{pages}{17953--19979}.
\bibitem[{Kresse and Joubert(1999)}]{Kresse1999}
\bibinfo{author}{Kresse\xfnm[ G.]}, \bibinfo{author}{Joubert\xfnm[ D.]}.
\newblock \bibinfo{journal}{Phys Rev B}
  \bibinfo{year}{1999};\bibinfo{volume}{59}:\bibinfo{pages}{1758--1775}.
\bibitem[{Perdew and Wang(1992)}]{Perdew92}
\bibinfo{author}{Perdew\xfnm[ J.P.]}, \bibinfo{author}{Wang\xfnm[ Y.]}.
\newblock \bibinfo{journal}{Phys Rev B}
  \bibinfo{year}{1992};\bibinfo{volume}{45}(\bibinfo{number}{23}):\bibinfo{pag%
es}{13244--13249}.
\bibitem[{Sinclair et~al.(1978)Sinclair, Gehlen, Hoagland and
  Hirth}]{Sinclair1978}
\bibinfo{author}{Sinclair\xfnm[ J.E.]}, \bibinfo{author}{Gehlen\xfnm[ P.C.]},
  \bibinfo{author}{Hoagland\xfnm[ R.G.]}, \bibinfo{author}{Hirth\xfnm[ J.P.]}.
\newblock \bibinfo{journal}{J Appl Phys}
  \bibinfo{year}{1978};\bibinfo{volume}{49}:\bibinfo{pages}{3890--3897}.
\bibitem[{Rao et~al.(1998)Rao, Hernandez, Simmons, Parthasarathy and
  Woodward}]{Rao1998}
\bibinfo{author}{Rao\xfnm[ S.]}, \bibinfo{author}{Hernandez\xfnm[ C.]},
  \bibinfo{author}{Simmons\xfnm[ J.P.]}, \bibinfo{author}{Parthasarathy\xfnm[
  T.A.]}, \bibinfo{author}{Woodward\xfnm[ C.]}.
\newblock \bibinfo{journal}{Philos Mag A}
  \bibinfo{year}{1998};\bibinfo{volume}{77}:\bibinfo{pages}{231--256}.
\newblock \doi{\bibinfo{doi}{10.1080/01418619808214240}}.
\bibitem[{Trinkle(2008)}]{TrinkleLGF2008}
\bibinfo{author}{Trinkle\xfnm[ D.R.]}.
\newblock \bibinfo{journal}{Phys Rev B}
  \bibinfo{year}{2008};\bibinfo{volume}{78}:\bibinfo{pages}{014110}.
\newblock \doi{\bibinfo{doi}{10.1103/PhysRevB.78.014110}}.
\bibitem[{Yasi et~al.(2010)Yasi, {Hector, Jr.} and Trinkle}]{Yasi2010}
\bibinfo{author}{Yasi\xfnm[ J.A.]}, \bibinfo{author}{{Hector, Jr.}\xfnm[
  L.G.]}, \bibinfo{author}{Trinkle\xfnm[ D.R.]}.
\newblock \bibinfo{journal}{Acta mater}
  \bibinfo{year}{2010};\bibinfo{volume}{58}:\bibinfo{pages}{5704--5713}.
\newblock \doi{\bibinfo{doi}{10.1016/j.actamat.2010.06.045}}.
\bibitem[{Yasi et~al.(2009)Yasi, Nogaret, Trinkle, Qi, {Hector, Jr.} and
  Curtin}]{Yasi2009}
\bibinfo{author}{Yasi\xfnm[ J.A.]}, \bibinfo{author}{Nogaret\xfnm[ T.]},
  \bibinfo{author}{Trinkle\xfnm[ D.R.]}, \bibinfo{author}{Qi\xfnm[ Y.]},
  \bibinfo{author}{{Hector, Jr.}\xfnm[ L.G.]}, \bibinfo{author}{Curtin\xfnm[
  W.A.]}.
\newblock \bibinfo{journal}{Model Simul Mater Sci Eng}
  \bibinfo{year}{2009};\bibinfo{volume}{17}:\bibinfo{pages}{055012}.
\newblock \doi{\bibinfo{doi}{10.1088/0965-0393/17/5/055012}}.
\bibitem[{Nogaret et~al.(2010)Nogaret, Curtin, Yasi, {Hector, Jr.} and
  Trinkle}]{Nogaret2010}
\bibinfo{author}{Nogaret\xfnm[ T.]}, \bibinfo{author}{Curtin\xfnm[ W.A.]},
  \bibinfo{author}{Yasi\xfnm[ J.A.]}, \bibinfo{author}{{Hector, Jr.}\xfnm[
  L.G.]}, \bibinfo{author}{Trinkle\xfnm[ D.R.]}.
\newblock \bibinfo{journal}{Acta mater}
  \bibinfo{year}{2010};\bibinfo{volume}{58}:\bibinfo{pages}{4332--4343}.
\newblock \doi{\bibinfo{doi}{10.1016/j.actamat.2010.04.022}}.
\bibitem[{{Ward Flynn} et~al.(1961){Ward Flynn}, Mote and Dorn}]{WardFlynn1961}
\bibinfo{author}{{Ward Flynn}\xfnm[ P.W.]}, \bibinfo{author}{Mote\xfnm[ J.]},
  \bibinfo{author}{Dorn\xfnm[ J.E.]}.
\newblock \bibinfo{journal}{Trans Metall Soc AIME}
  \bibinfo{year}{1961};\bibinfo{volume}{221}:\bibinfo{pages}{1148--1154}.
\bibitem[{Kocks et~al.(1975)Kocks, Argon and Ashby}]{Kocks1975}
\bibinfo{author}{Kocks\xfnm[ U.F.]}, \bibinfo{author}{Argon\xfnm[ A.S.]},
  \bibinfo{author}{Ashby\xfnm[ M.F.]}.
\newblock \bibinfo{journal}{Prog Mater Sci}
  \bibinfo{year}{1975};\bibinfo{volume}{19}:\bibinfo{pages}{1--291}.
\bibitem[{P{\"u}schl(2002)}]{Pueschl2002}
\bibinfo{author}{P{\"u}schl\xfnm[ W.]}.
\newblock \bibinfo{journal}{Prog Mater Sci}
  \bibinfo{year}{2002};\bibinfo{volume}{47}:\bibinfo{pages}{415--461}.
\bibitem[{Argon(2006)}]{Argon2005}
\bibinfo{author}{Argon\xfnm[ A.S.]}.
\newblock \bibinfo{title}{Strengthening Mechanisms in Crystal Plasticity}.
\newblock \bibinfo{publisher}{Oxford University Press}; \bibinfo{year}{2006}.
\bibitem[{Humphrey et~al.(1996)Humphrey, Dalke and Schulten}]{Humphrey1996}
\bibinfo{author}{Humphrey\xfnm[ W.]}, \bibinfo{author}{Dalke\xfnm[ A.]},
  \bibinfo{author}{Schulten\xfnm[ K.]}.
\newblock \bibinfo{journal}{Journal of Molecular Graphics}
  \bibinfo{year}{1996};\bibinfo{volume}{14}:\bibinfo{pages}{33--38}.

\end{thebibliography}
\end{document}